\documentclass{aa}      
\usepackage[varg]{txfonts}
\usepackage{geometry}                           
\usepackage{graphicx}                           
\usepackage{amssymb}
\usepackage{color}

\title{Congruity of the Crab Pulsar's gamma-ray spectrum with the spectral distribution of tightly focused caustics}
\author{Houshang Ardavan}
\institute{Institute of Astronomy, University of Cambridge, Madingley Road, Cambridge CB3 0HA, UK\\
email: ardavan@ast.cam.ac.uk}
\date{}                                                 

\begin{document}

\abstract
{The spectrum derived here for the most tightly focused component of the radiation generated by the super-luminally moving current sheet in the magnetosphere of a non-aligned neutron star has a distribution function that fits the entire gamma-ray spectrum of the Crab Pulsar on its own.  This is the first time that the undivided breadth of this spectrum, from $10^2$ to $10^6$ MeV, is not only described by a single distribution function but is also explained by means of a single emission mechanism.  To illustrate that the derived function describes the spectral distribution of the high-energy emission from any non-aligned neutron star, we analyse, in addition, the spectra of two other gamma-ray pulsars for which sufficiently large datasets are available: PSR J0101-6422 and PSR J1709-4429.  From the connection between the parameters of the fitted spectra and the physical characteristics of their sources, we moreover infer certain attributes of the magnetospheres of the analysed pulsars: the angle between the magnetic and spin axes of their central neutron stars, the scale factor of the electric current density that is associated with their current sheet, and the privileged latitudinal direction (relative to the spin axis) in which they are observed.}


  \maketitle

\section{Introduction}
\label{sec:introduction}
Numerical computations based on the force-free and particle-in-cell formalisms have now firmly established that the magnetosphere of a non-aligned neutron star entails a current sheet outside its light cylinder whose rotating distribution pattern moves with linear speeds exceeding the speed of light in a vacuum~\citep[see the review article][and the references therein]{Philippov2022}.  A study of the characteristics of the radiation that is generated by this super-luminally moving current sheet has in turn provided an all-encompassing explanation for the salient features of the radiation received from pulsars: its brightness temperature, polarisation, spectrum, and profile with microstructure and with a phase lag between the radio and gamma-ray peaks~\citep{Ardavan2021, Ardavan2022a}.  

The radiation field generated by a constituent volume element of the current sheet in the magnetosphere of a neutron star features a synergy between the super-luminal version of the field of synchrotron radiation and the vacuum version of the field of \v{C}erenkov radiation.  Once superposed to yield the emission from the entire volume of the source, the contributions from the volume elements of the current sheet that approach the observation point at the speed of light and with zero acceleration at the retarded time interfere constructively and form caustics in certain latitudinal directions relative to the spin axis of the neutron star.  The waves that embody these caustics are more focused the farther they are from their source: as their distance from their source increases, two nearby stationary points of their phases draw closer to each other and eventually coalesce at infinity~\citep[][Sect. 4]{Ardavan2021}.  By virtue of their extremely narrow peaks in the time domain, such focused pulses have broad spectra that encompass X-ray and gamma-ray frequencies~\citep[][Table~1 and Sect.~5.3]{Ardavan2021}.  In directions where the stationary points of their phases are isolated, the emitted pulses are not focused as strongly and so are detectable at lower frequencies~\citep[][Sect.~5.4]{Ardavan2021}.

A heuristic account of the radiation that is generated by the current sheet in the magnetosphere of a neutron star is outlined in Sect.~\ref{sec:heuristic}.  The contribution towards the spectrum of this radiation from the vicinity of its tightly focused caustics is derived in Sect.~\ref{sec:spectrum}, and it is shown that the derived spectral distribution function fits the observed gamma-ray spectrum of the Crab Pulsar~\citep{Abdo2010, Aleksic2011,Ansoldi2016} over its entire breadth, from $10^2$ to $10^6$ MeV (Sect.~\ref{sec:fit}).  The same analysis is applied to the data on the spectra of two other gamma-ray pulsars, PSR J0101-6422 and PSR J1709-4429~\citep{Abdo2013}, in Sect.~\ref{sec:examples}, and the parameters of the observed spectra of the three pulsars in question are related to the physical characteristics of their sources in Appendix~\ref{App}.  The contrast between the single radiation mechanism underlying the present spectral distribution function and the disparate radiation mechanisms and spectral distribution functions currently invoked in the literature for fitting the data in different sections of the Crab Pulsar's gamma-ray spectrum~\citep[the review articles][and the references therein]{Zanin2017,Amato2021} is briefly commented on in Sect.~\ref{sec:conclusion}.

\section{A heuristic account of the radiation from the super-luminally moving current sheet in the magnetosphere of a neutron star}
\label{sec:heuristic}

Results of the mathematical treatment of the radiation by the super-luminally moving current sheet in the magnetosphere of a neutron star, which was presented in~\citet{Ardavan2021}, are explained here in more transparent physical terms with the aid of illustrations.

\begin{figure}
\centerline{\includegraphics[width=17cm]{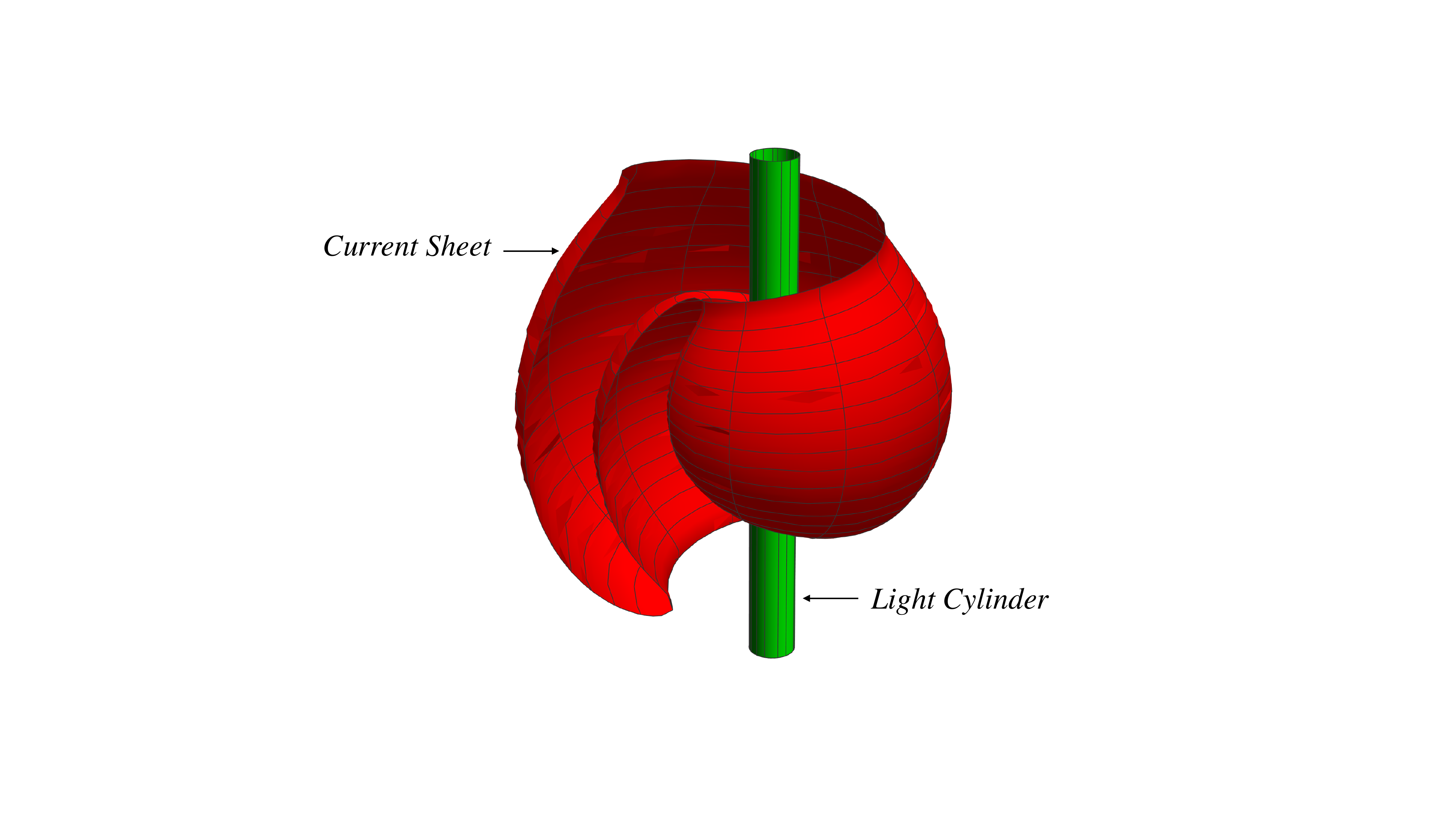}}
\caption{Snapshot of a single turn of the current sheet about the light cylinder (i.e.\ $r=c/\omega$, where $r$ is the distance from the rotation axis, $\omega$ is the angular frequency of rotation of the central neutron star, and $c$ is the speed of light in a vacuum) for the value $\alpha=\pi/3$ of the angle between the magnetic and rotation axes of the star.  This surface undulates within the latitudinal interval $\pi/2-\alpha\le\theta\le\pi/2+\alpha$ each time it turns about the rotation axis, thus wrapping itself around the light cylinder as it extends to the outer edge of the magnetosphere.}
\label{FH0}
\end{figure}

The surface on which the magnetospheric current sheet is distributed (Fig.~\ref{FH0}) spirals away from the light cylinder in the azimuthal direction at the same time as undulating in the latitudinal direction~\citep{Tchekhovskoy:etal, Bogovalov1999}.  Its motion consists of a rigid rotation with the same angular frequency, $\omega$, as that of the central neutron star and a radial expansion (resulting from the combination of its spiral structure and rigid rotation) at the speed of light in a vacuum, $c$.  This is not incompatible with the requirements of special relativity because the super-luminally moving distribution pattern of the current sheet is created by the coordinated motion of aggregates of sub-luminally moving charged particles~\citep{BolotovskiiBM:VaveaD, GinzburgVL:vaveaa, BolotovskiiBM:Radbcm}.

In~\citet{Ardavan2021}, the distribution of charges and currents that make up the current sheet at any given time is treated as a prescribed volume source whose density can be inserted in the retarded solution of the inhomogeneous Maxwell's equations to find the radiation field it generates in unbounded free space.  The only role assigned to the rest of the magnetosphere, whose radiation field is negligibly weaker than that of the current sheet, is to maintain the propagation of this sheet.  The multi-wavelength focused pulses emitted by the current sheet escape the plasma surrounding the neutron star in the same way that the radiation generated by the accelerating charged particles invoked in most current attempts at modelling the emission mechanism of these objects does~\citep{Philippov2018,Philippov2019}.  

\begin{figure*}
\centerline{\includegraphics[width=17cm]{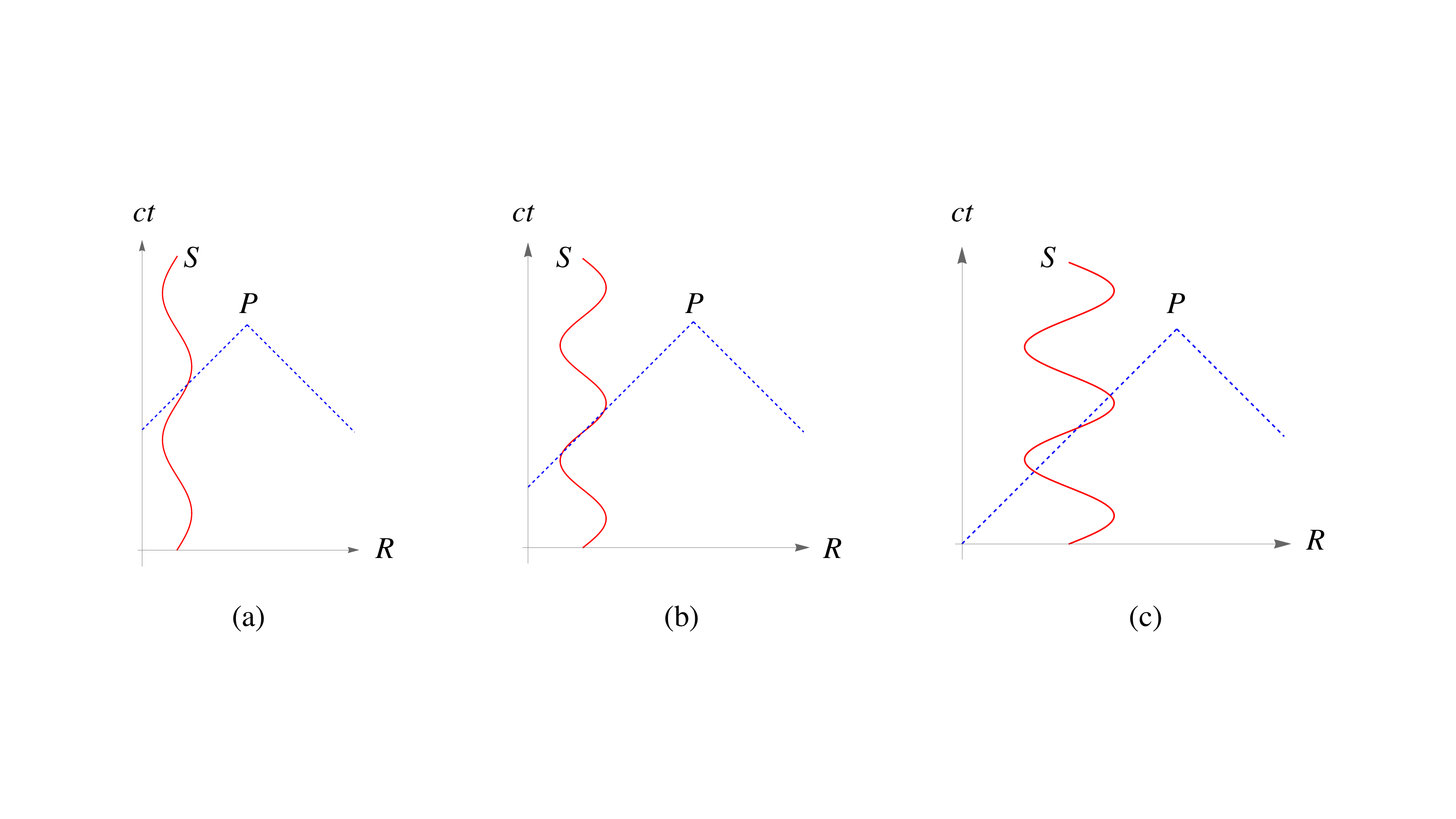}}
\caption{Space-time depictions of the past light cone of the observation point, $P,$ located at ${\bf x}_P$ (the dashed blue  lines) and the possible forms of the trajectory, $R(t)=\vert{\bf x}(t)-{\bf x}_P\vert$, of a super-luminally rotating source point, $S,$ with the position vector ${\bf x}$ (the red curves).  In (a), where $\vert{\rm d}R/{\rm d}t\vert<c$, the trajectory of $S$ intersects the past light cone of any $P$ only once, so the field observed at $P$ at time $t_P$ is determined by the single wave front that was emitted by $S$ at the retarded time, $t$.  In (c), where the component of the velocity of $S$ along the radiation direction is greater than $c$, the trajectory of $S$ intersects the past light cone of certain observers, $P$, at three points, so the field at $P$ receives contributions from three distinct values of the retarded time simultaneously. In (b), where both ${\rm d}R/{\rm d}t=-c$ and ${\rm d}^2R/{\rm d}t^2=0$ at the point where the trajectory of $S$ intersects the past light cone of $P$ tangentially, the field is determined by the coalescence of three contributions originating from a retarded time at which the source point approaches the observer along the radiation direction not only at the speed of light but also with zero acceleration.}
\label{FH1}
\end{figure*}

The current sheet is described by charge and current densities whose space-time distributions depend on the azimuthal coordinate $\varphi$ and time $t$ in the combination $\varphi-\omega t$ only~\citep{Tchekhovskoy:etal}.  The radiation field we are after can be built up, therefore, by the superposition of the fields of the uniformly rotating volume elements that constitute this source.  The radiation field of a rotating volume element whose linear speed exceeds the speed of light in a vacuum receives simultaneous contributions from more than one retarded position of the source at certain observation points (Fig.~\ref{FH1}).  

As a result, this field entails intersecting wave fronts that possess a two-sheeted cusped envelope (Figs.~\ref{FH2} and \ref{FH3}).  Outside the envelope only one wave front passes through the observation point at any given observation time; but inside the envelope three distinct wave fronts, emitted at three different values of the retarded time, simultaneously pass through each observation point (Sect.~\ref{sec:PointSource}).  Coalescence of two of the contributing retarded times on the envelope of wave fronts results in the divergence of the resulting field on this surface (reflecting the fact that a super-luminally moving source cannot be point-like; \citealt{BolotovskiiBM:VaveaD, GinzburgVL:vaveaa, BolotovskiiBM:Radbcm}).  At an observation point on the cusp locus of the envelope all three of the contributing retarded times coalesce and the field in question has a higher-order singularity.  This radiation field embraces a synergy between the vacuum version of the field of {\v C}erenkov radiation and the super-luminal version of the field of synchrotron radiation. 

Constructive interference of the emitted waves and formation of caustics thus play a crucial role in determining the radiation field of the current sheet (Sect.~\ref{sec:volume}).  Its double-peaked pulse profiles and S-shaped polarisation position angle distributions stem from the caustics associated with the nearby stationary points of certain phase functions~\citep[][Sect. 5.1]{Ardavan2021}.   What underpin the high brightness temperatures and the broad frequency spectra of this radiation are the extraordinary values of the amplitude and width of the pulses that are generated when the maxima and minima of the phase functions in question coalesce into inflection points~\citep[][Sects. 5.2 and 5.3]{Ardavan2021}.  There is always a latitudinal direction along which the separation between any pair of these nearby maxima and minima decreases with increasing distance from the source.  The enhanced focusing of the emitted waves that takes place as a result of the diminishing separation between the nearly coincident stationary points of a phase function gives rise, in turn, to a lower rate of decay of the flux density of the radiation with distance~\citep[][Sect. 5.5]{Ardavan2021}.  That the decay of the present radiation with distance disobeys the inverse-square law in certain directions~\citep[see also][]{Ardavan2023} is not incompatible with the requirements of the conservation of energy because the radiation process discussed here is intrinsically transient: the difference in the fluxes of power across any two spheres centred on the star is balanced by the change with time of the energy contained inside the shell bounded by those spheres~\citep[see][Appendix C, where this is demonstrated for each high-frequency Fourier component of a super-luminally rotating source distribution]{Ardavan2019}.  

\subsection{The field generated by a constituent volume element of the source}
\label{sec:PointSource}

A super-luminal source is necessarily volume-distributed~\citep{BolotovskiiBM:VaveaD, GinzburgVL:vaveaa, BolotovskiiBM:Radbcm}.  However, its field can be built up from the superposition of the fields of its constituent volume elements, which are point-like.  The current sheet in the magnetosphere of a non-aligned neutron star that rotates with the angular frequency $\omega$ is described by charge and current densities outside the light cylinder whose distribution patterns rigidly rotate with the same angular frequency as that of the central neutron star~\citep{Tchekhovskoy:etal}.  The radiation field that is generated by this current sheet can be built up, therefore, by the superposition of the fields of uniformly rotating volume elements whose linear speeds $r\omega$ exceed $c$, where $r$ is the distance of a source point from the rotation axis.  

In the space-time $({\bf x}, t)$ of source points, the past light cone $\vert{\bf x}-{\bf x}_P\vert=c(t_P-t)$ of an observer, $P$, with the space-time coordinates $({\bf x}_P, t_P)$ is a sphere centred on the observer's position, ${\bf x}_P$, which collapses, at the speed of light, $c$, onto the point ${\bf x}_P$ at the observation time $t_P$.  Since the waves it emits propagate with the speed $c$, a moving source point ${\bf x}={\bf x}(t)$ makes its contribution towards the radiation field detected at $({\bf x}_P, t_P)$ at the retarded (emission) time $t$ when this collapsing sphere crosses it, that is, at the instant at which $R(t)\equiv\vert{\bf x}(t)-{\bf x}_P\vert=c(t_P-t)$.  The red curves in Fig.~\ref{FH1} depict the possible trajectories of a rotating source point in the $(R, ct)$ space.  The dashed lines in this figure represent the past light cone of the observer $P$.  Depending on their positions relative to the observer, various source points (volume elements) of a super-luminally rotating source distribution approach the observer along the radiation direction $({\bf x}_P-{\bf x})/\vert{\bf x}_P-{\bf x}\vert$ with a speed ${\rm d}R/{\rm d}t$ at the retarded time (i.e.\ the time at which the space-time trajectory of the source element in question intersects the past light cone of the observer) that can be less than, equal to, or greater than $c,$ as in Figs.~\ref{FH1}a--\ref{FH1}c.  

The number of times the collapsing sphere $\vert{\bf x}-{\bf x}_P\vert=c(t_P-t)$ intersects a rotating source point at ${\bf x}(t)$ during the time it propagates past the circular orbit of such a point source depends on whether $\vert{\rm d}R/{\rm d}t\vert$ is less than, equal to, or greater than $c$.  The trajectory of a source element for which $\vert{\rm d}R/{\rm d}t\vert<c$ intersects the past light cone of any observer only once and so makes its contribution towards the value of the observed radiation field at a single instant of retarded time (Fig.~\ref{FH1}a).  In contrast, a source element for which $\vert{\rm d}R/{\rm d}t\vert>c$ makes its contribution towards the value of the radiation field detected by certain observers at three distinct instants of the retarded time: a contribution comprising three distinct parts that are received simultaneously at the observation time $t_P$ (Fig.~\ref{FH1}c).  (For a sufficiently large $\vert{\rm d}R/{\rm d}t\vert$, the trajectory of the rotating source point can intersect the past light cone of the observer 5, 7, 9, \ldots times.)  The three parts of this contribution coalesce in the case of a source element that approaches the observer at the speed of light (${\rm d}R/{\rm d}t=-c$) and with zero acceleration (${\rm d}^2R/{\rm d}t^2=0$) at the retarded time (Fig.~\ref{FH1}b). 

\begin{figure*}
\centerline{\includegraphics[width=17cm]{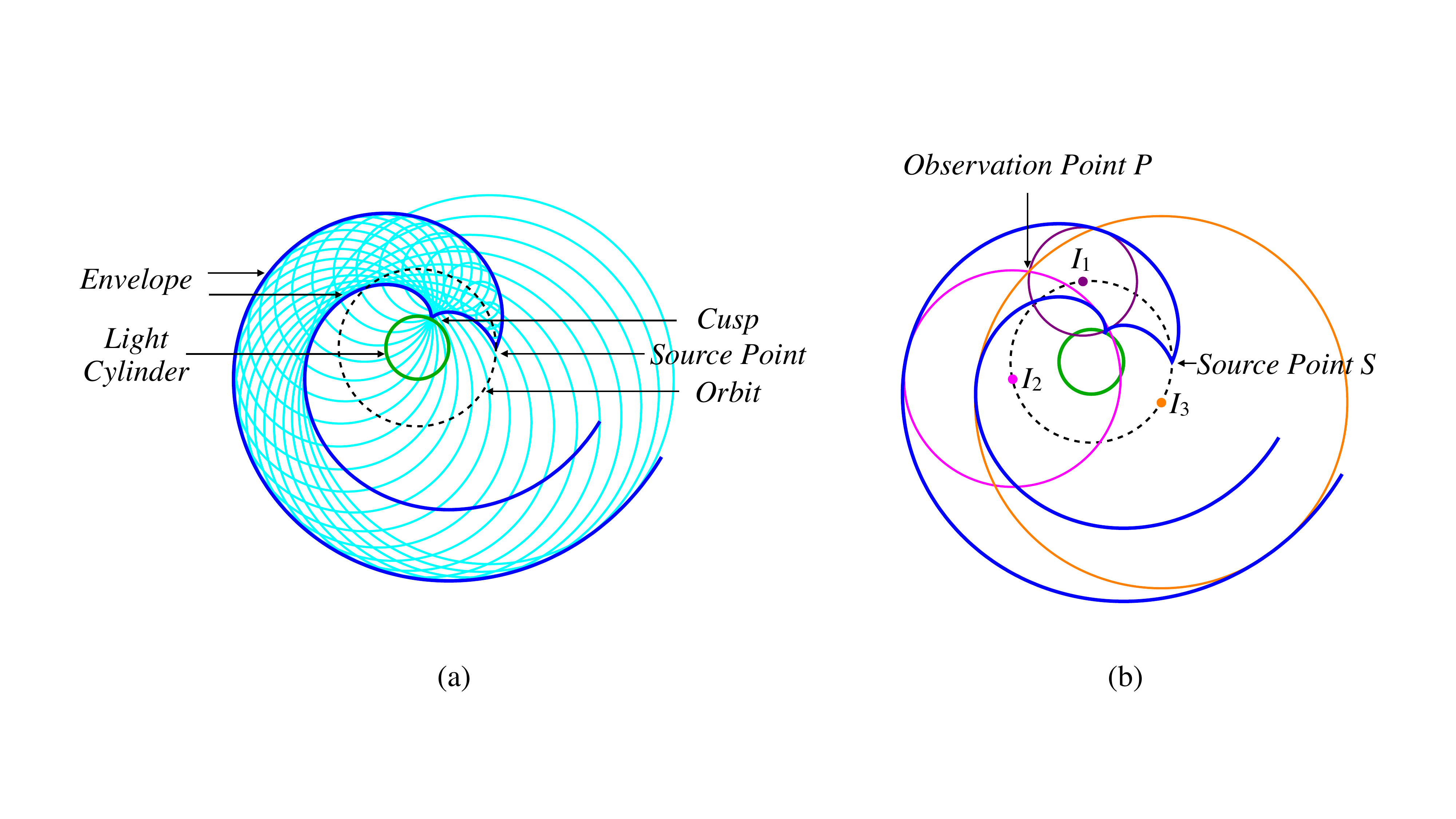}}
\caption{Wave fronts emanating from a uniformly rotating source point whose linear speed exceeds the speed of light, $c$.  (a) Cross-sections of the wave fronts (the circles in light blue) emanating from a volume element of the current sheet with fixed radial and latitudinal coordinates.  This figure is plotted for a source element the radius of whose orbit (the dotted circle) is $2.5$ times the radius $c/\omega$ of the light cylinder (the green circle).  Cross-sections of the two sheets of the envelope of these wave fronts with the plane of the orbit (shown in dark blue) meet at a cusp and wind around the rotation axis, while moving away from it all the way to the far zone. (b) By receiving three wave fronts (the circles in brown, pink, and orange) centred at three distinct retarded positions ($I_1$, $I_2$, and $I_3$) of the source point $S$, an observer, $P$, inside the envelope detects three images of $S$ simultaneously  ($I_1$, $I_2$,  and $I_3$).} 
\label{FH2}
\end{figure*}

Figure~\ref{FH2}a shows the intersections with the equatorial plane of the spherical wave fronts emitted by a source element rotating with the constant angular frequency $\omega$ whose orbit lies outside the light cylinder $r_P=c/\omega$ (i.e.\ whose speed exceeds $c$), where $r_P$ is the distance of the observation point from the rotation axis.  The envelope of these wave fronts consists of two sheets that meet at a cusp.  Only a single wave front propagates past an observer outside the envelope (for whom $\vert{\rm d}R/{\rm d}t\vert<c$) at any given time.  But an observer inside the envelope (for whom $\vert{\rm d}R/{\rm d}t\vert>c$) receives three intersecting wave fronts, emanating from three distinct retarded positions of the source element ($I_1$, $I_2$ and $I_3$), simultaneously (see Fig.~\ref{FH2}b).  On each sheet of the envelope, two of these wave fronts (together with the corresponding retarded times at which they are emitted) coalesce and interfere constructively to generate an infinitely large radiation field~\citep[][Sect. 3.4]{Ardavan2021}.  On the cusp locus of the envelope, where all three of these wave fronts (together with the corresponding images $I_1$, $I_2$ and $I_3$ of the source element) coalesce, the resulting radiation field has a higher-order singularity.  (Divergence of the field on the envelope and its cusp, which stem from the relativistic restrictions inherent in electrodynamics, reflect the fact that no super-luminal source can be point-like; \citealt{BolotovskiiBM:VaveaD, GinzburgVL:vaveaa, BolotovskiiBM:Radbcm}.)

\begin{figure*}
\centerline{\includegraphics[width=17cm]{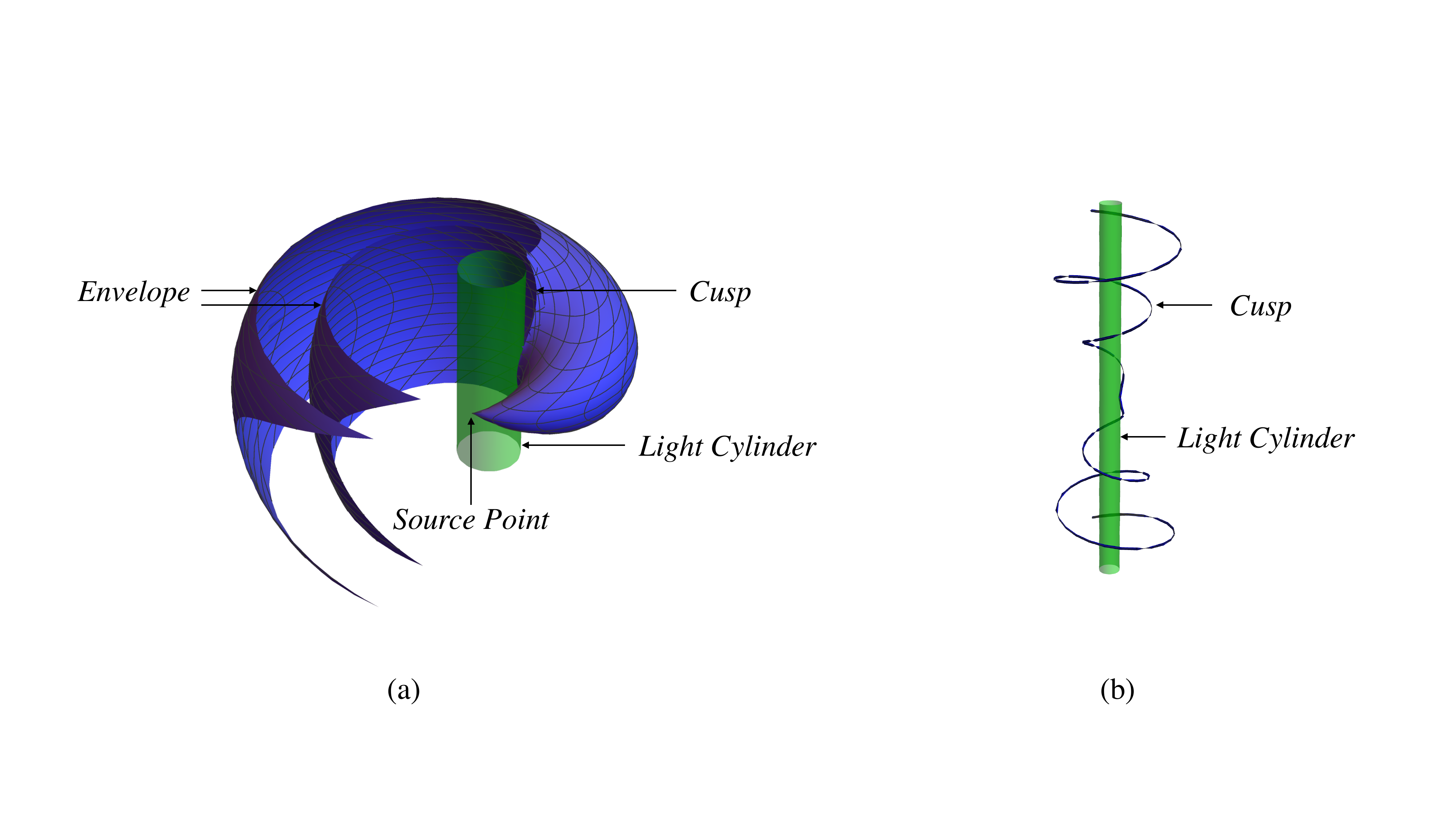}}
\caption{Cusped envelope of the wave fronts emanating from a uniformly rotating source point whose linear speed exceeds the speed of light, $c$.  (a) Three-dimensional view of the two-sheeted envelope of wave fronts emanating from a volume element of the current sheet.  This envelope spirals into the far zone, maintaining a symmetry with respect to the plane of the orbit of the source element.  (b) Cusp locus along which the two sheets of the envelope meet.  This curve tangentially touches the light cylinder where it crosses the plane of the orbit and spirals into the far zone on a hyperboloid whose asymptotes lie on the cones $\theta_P=\arcsin[c/(r_{\rm S}\omega)]$ and $\theta_P=\pi-\arcsin[c/(r_{\rm S}\omega)]$, where $r_{\rm S}$ is the radius of the circular orbit of the source element.}
\label{FH3}
\end{figure*}

Figure~\ref{FH3}a shows a three-dimensional view of the envelope of wave fronts that are emitted by a uniformly rotating source element outside the light cylinder.  The cusp along which the two sheets of this envelope meet touches the light cylinder on the plane of the orbit tangentially and spirals away from the rotation axis on a hyperboloid that is symmetrical with respect to that plane (see Fig.~\ref{FH3}b).  As in Fig.~\ref{FH2}, the number of wave fronts that pass the observation point at a given observation time are one, two or three outside, on or inside the envelope.  Likewise, the generated radiation is most tightly focused on the cusp curve shown in Fig.~\ref{FH3}b where the waves from three coalescent retarded positions (images) of the source element interfere constructively.

\begin{figure*}
\centerline{\includegraphics[width=17cm]{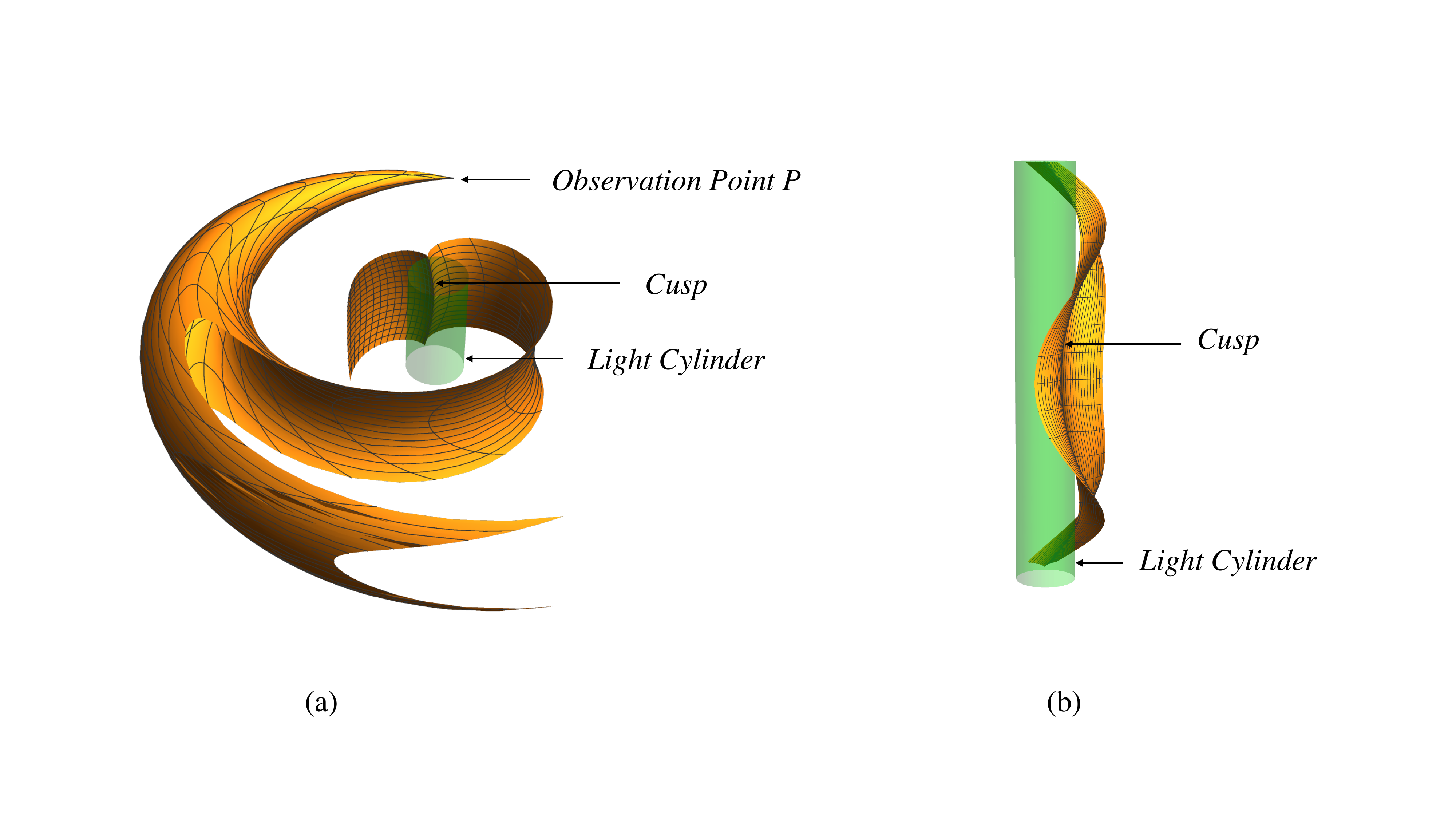}}
\caption{Bifurcation surface of the observation point, $P$, its cusp locus, and the light cylinder $r=c/\omega$.  (a) In contrast to the envelope of wave fronts shown in Fig.~\ref{FH3}a, which resides in the space of observation points, the surface shown here resides in the space of source points: it is the locus of source elements that approach $P$ along the radiation direction at the speed of light at the retarded time.  The two sheets of this surface meet along a cusp that tangentially touches the light cylinder where it crosses the plane of rotation containing the observation point and spirals outwards as the colatitude, $\theta$, decreases.  The source points on this cusp approach the observer along the radiation direction not only at the speed of light but also with zero acceleration at the retarded time.  (b) Close-up view of the two sheets of the bifurcation surface in the vicinity of the cusp locus of this surface.  The hyperboloid containing this cusp has the conical asymptotes $\theta=\arcsin[c/(r_P\omega)]$ and $\theta=\pi-\arcsin[c/(r_P\omega)]$.}
\label{FH4}
\end{figure*} 

\subsection{The field generated by the entire volume of the source}
\label{sec:volume}

Depending on the speed $\vert{\rm d}R/{\rm d}t\vert$ with which they approach an observation point $P$ at the retarded time, different volume elements of an extended source make differing contributions towards the value of the radiation field at $P$ (see Fig.~\ref{FH1}).  To distinguish between the contributions from different sections of a volume source, we need to introduce the notion of bifurcation surface: a two-sheeted cusped surface reciprocal to the envelope of wave fronts that resides in the space of source points, instead of residing in the space of observation points, and issues from the observation point, instead of issuing from a source point (see Fig.~\ref{FH4}).  

The intersection of the bifurcation surface of an observation point with the volume of the source divides this volume into two parts.  The source elements inside the bifurcation surface make their contributions towards the observed field at three distinct values of the retarded time, while the source elements outside the bifurcation surface make their contributions at a single value of the retarded time (as a sub-luminally moving source would).  The source elements inside and close to the bifurcation surface (for which the values of two of the contributing retarded times approach one another) and the source elements inside the bifurcation surface close to its cusp (for which all three values of the contributing retarded times coalesce) are by far the dominant contributors towards the strength of the observed field.  

As noted before, the densities of charges and currents in the magnetosphere of a non-aligned neutron star that rotates with the constant angular frequency $\omega$ are functions of the azimuthal angle $\varphi$ and time $t$ in the combination ${\hat\varphi}=\varphi-\omega t$ only~\citep{Tchekhovskoy:etal}.  The coordinate ${\hat\varphi}$ labels each volume element of the uniformly rotating magnetosphere with its azimuthal position at the time $t=0$ (thus ranging over an interval of length $2\pi$).  To superpose the contributions from various volume elements of such a magnetosphere towards the value of the radiation field, we must perform the integration in the classical expression for the retarded field~\citep[][Sect. 3.2]{Ardavan2021} over the volume occupied by the charges and currents in the $(r_{\rm s},{\hat\varphi}, \theta)$ space, where the radius, $r_{\rm s}$, and the colatitude, $\theta$, denote the corresponding spherical coordinates of these source elements.

\begin{figure*}
\centerline{\includegraphics[width=17cm]{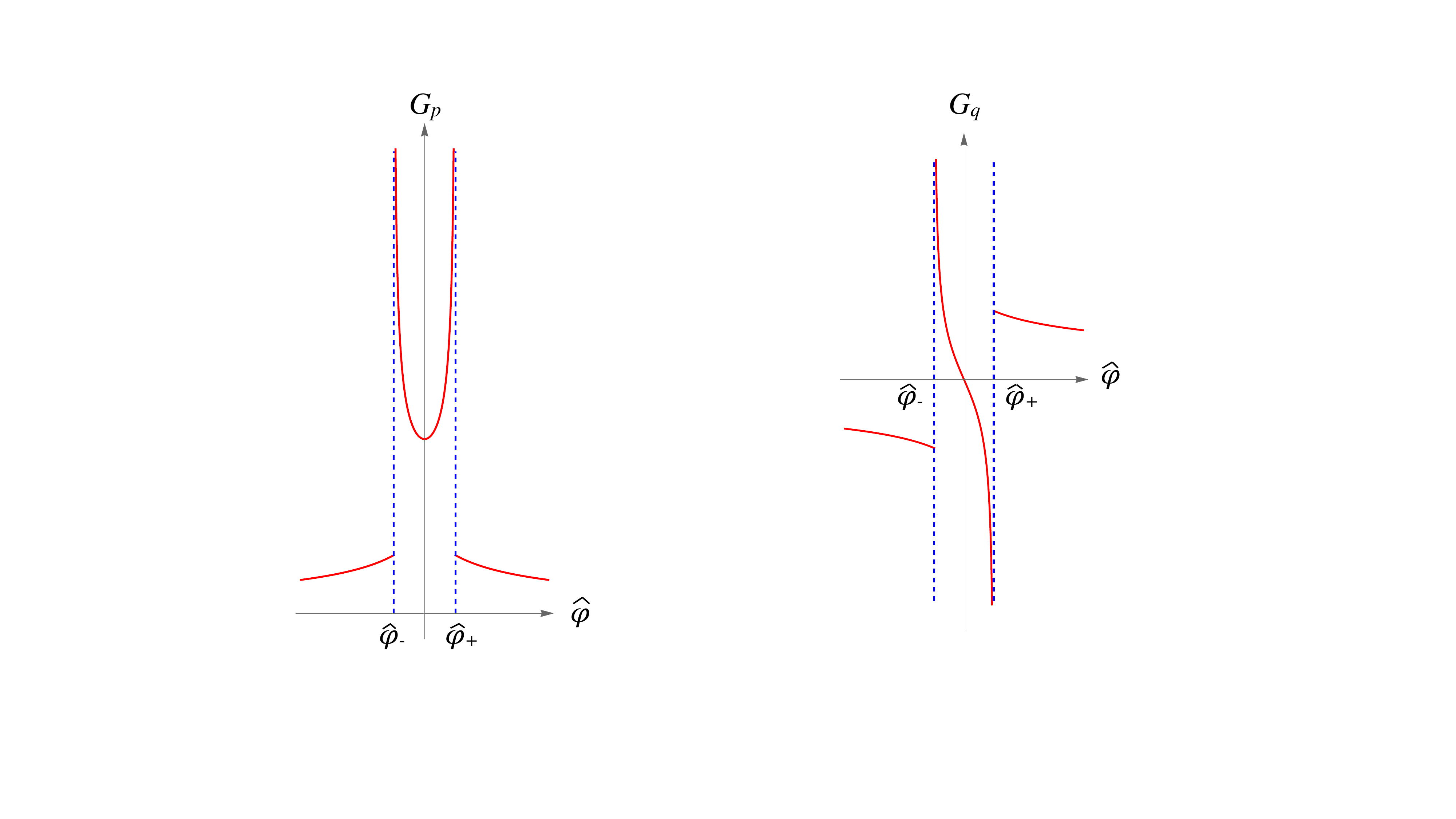}}
\caption{Schematic depiction of the ${\hat\varphi}$ dependence of the functions that describe the contribution, towards the observed field, of a super-luminally rotating source element inside (${\hat\varphi}_-<{\hat\varphi}<{\hat\varphi}_+$) and outside (${\varphi}\le{\hat\varphi}_-$, ${\varphi}\ge{\hat\varphi}_+$) a cusped strip of the bifurcation surface (Fig.~\ref{FH4}b).  The contribution in question is described by a linear combination of these two functions.}
\label{FH5}
\end{figure*}

By virtue of approaching the observer with a speed that is close to $c$ and an acceleration that is vanishingly small at the retarded time, the source elements in the vicinity of the strip of the bifurcation surface that borders on its cusp (Fig.~\ref{FH4}b) make a contribution towards the observed radiation field that is singular on the inner and regular on the outer side of the bifurcation surface: the ${\hat\varphi}$ dependence of the contributions made by the elements inside (${\hat\varphi}_-<{\hat\varphi}<{\hat\varphi}_+$) and outside (${\hat\varphi}\le{\hat\varphi}_-$, ${\hat\varphi}\ge{\hat\varphi}_+$) a cusped strip of the bifurcation surface (Fig.~\ref{FH4}b) is given by a linear combination of the two functions shown in Fig.~\ref{FH5}.  The contribution of each constituent ring ($r_{\rm s}=$ const., $\theta=$ const., $0\le{\hat\varphi}<2\pi$) of a super-luminally rotating source distribution towards the field detected at an observation point $P$ is thus discontinuous across the two sheets of the bifurcation surface associated with $P$~\citep[][Sect. 3.3]{Ardavan2021}.

Superposition of the contributions of the source elements that constitute a rotating ring ($r_{\rm s}=$ const., $\theta=$ const., $0\le{\hat\varphi}<2\pi$) towards the field (i.e.\ an integration of the classical expression for the retarded field with respect to ${\hat\varphi}$ at fixed values of $r_{\rm s}$ and $\theta$) shows that, in addition to a contribution identical to the conventional one, which is obtained in the case of a sub-luminal source, the present radiation field receives contributions from the discontinuities (depicted in Fig.~\ref{FH5}) across the bifurcation surface of the observation point~\citep[][Sect. 3.5]{Ardavan2021}.  These additional contributions, as we shall see below, turn out to result in an unconventional radiation field, much stronger than that resulting from the conventional contribution, whose characteristics differ from those of any previously known radiation field~\citep[][Sect. 4.1]{Ardavan2021}.

Next, we need to superpose the fields of the constituent rotating rings of the source distribution that lie on a cone outside the light cylinder, that is,\ the rings for which $\theta$ is constant and $r_s$ ranges from $c/(\omega\sin\theta)$ to the outer edge of the magnetosphere.  The dominant contribution towards the value of the field of this truncated cone at the observation point $P$ happens to come from those source elements of it that lie in the vicinity of the cusp locus of the bifurcation surface of $P$, namely,\ the elements that approach the observation point along the radiation direction with speeds that are close to $c$ and accelerations that are vanishingly small at the retarded time~\citep[][Sect. 4.2]{Ardavan2021}.  This is not an unexpected result, given that the waves that emanate from such elements are focused most tightly (see Sect.~\ref{sec:PointSource}).

\begin{figure*}
\centerline{\includegraphics[width=17cm]{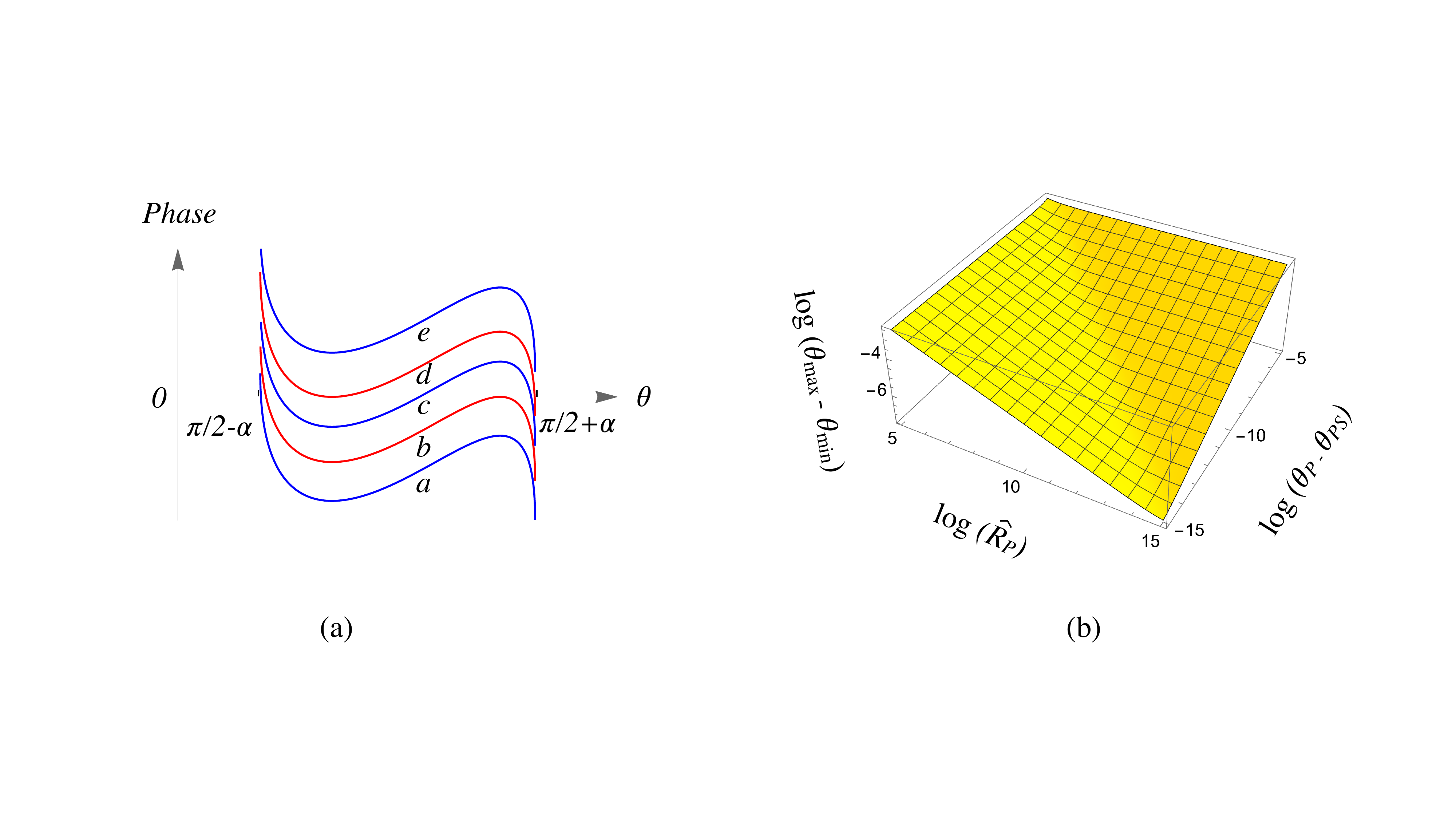}}
\caption{Phase of the kernel of the integral over $\theta$ and its critical points.  (a) Dependence on the colatitude, $\theta$, of the phase of the type of oscillating exponential that is encountered after we superpose the contributions of the constituent rotating rings of the source distribution that have the same colatitude but differing radii (see Sect.~\ref{sec:volume}).  The curves $a$--$e$ correspond to successively increasing values of the azimuthal coordinate of the observation point, $\varphi_P$.  Note that for curves $b$ and $d$ (shown in red), this phase vanishes at one of its turning points.  (b) Dependence of the separation between the locations of the maximum and minimum of a phase, $\theta_{\rm max}-\theta_{\rm min}$, with nearby turning points on the distance of the observer from the source,  ${\hat R}_P$, and on the departure, $\theta_P-\theta_{PS}$, of the colatitude of the observer, $\theta_P$, from the critical colatitude, $\theta_{PS}$, at which the maximum and minimum of such a phase coalesce.  This figure shows that $\theta_{\rm max}-\theta_{\rm min}$ decreases as ${\hat R}_P^{-1/2}$ with increasing distance when $(\theta_P-\theta_{PS})\lesssim{\hat R}_P^{-1}$.}
\label{FH6}
\end{figure*}

The resulting field of the aggregate of super-luminally rotating source elements that constitute the truncated cone in question is described by a linear combination of four oscillating exponentials whose phases are proportional to the radiation frequency.  As functions of the colatitude $\theta$ of the source elements, the phases of these exponential factors either have two turning points (as shown in Fig.~\ref{FH6}a) or vary monotonically~\citep[][Sect. 4.4]{Ardavan2021}.  For radiation frequencies that are appreciably larger than the rotation frequency $\omega/2\pi$, therefore, the dominant contribution towards the value of the field of the entire source (i.e.\ the field that results from the superposition of the fields of the truncated cones with differing opening angles, $\theta$) is expected to arise from the neighbourhoods of those values of $\theta$ at which the phases of one or more of these oscillating exponentials have turning points: the rapid oscillations (between positive and negative values) of these exponential factors as functions of $\theta$ bring about the cancellation of the contributions from all other values of $\theta$ when the fields of the truncated cones that make up the entire source are superposed~\citep[][Sect. 4.5]{Ardavan2021}.

Changes in the value of the azimuthal coordinate $\varphi_P$ of the observation point simply shift the curve representing the $\theta$ dependence of the phase of a given exponential factor up or down without altering its shape: the curves $a$--$e$ in Fig.~\ref{FH6}a correspond to successively increasing values of the longitude $\varphi_P$ across the window occupied by the emitted signal.  In the case of the phase function depicted in Fig.~\ref{FH6}a, the amplitude of the emitted signal peaks at the two longitudes at which a stationary point of this phase function crosses the horizontal axis, that is,\ at the values of $\varphi_P$ for which this phase function vanishes at its maximum or minimum (as do the curves labelled $b$ and $d$ in Fig.~\ref{FH6}a).  Depending on the fraction of the four oscillating exponentials whose phases have turning points, the number of peaks of the emitted signal can thus range from two to eight~\citep[][Sect. 4.4]{Ardavan2021}.

For any given inclination angle $\alpha$, there are a number (between two and eight) of critical values of the colatitude $\theta_P$ of the observation point (denoted as $\theta_{PS}$) at which the maximum and minimum of the phase of one or more of the exponentials that have two turning points coincide and form an inflection point.  Close to the critical colatitude, $\theta_{PS}$, the separation between the locations of the maximum and minimum of such phases,  $\theta_{\rm max}-\theta_{\rm min}$, has the dependence shown in Fig.~\ref{FH6}b on the departure of the colatitude of the observer from its critical value, $\theta_P-\theta_{PS}$, and on the distance of the observer from the source,  ${\hat R}_P$, in units of the light-cylinder radius.  It can be seen from Fig.~\ref{FH6}b that the separation between the maximum and minimum of the phase shown in Fig.~\ref{FH6}a decreases as ${\hat R}_P^{-1/2}$ with increasing distance of the observer from the source when $(\theta_P-\theta_{PS})\lesssim{\hat R}_P^{-1}$, a result that implies that the emitted signal gets more focused the farther away it is from its source~\citep[][Sect. 5.5]{Ardavan2021}. 

If we now attempt to integrate the resulting expression for the field of the entire source over all frequencies, to obtain the time-domain description of this field, we encounter a divergent integral.  This divergence stems from the fact that, by entailing a step-like discontinuity in the magnetic field~\citep{Tchekhovskoy:etal}, the numerical simulations of the magnetosphere assign a zero width to the current sheet.  The vanishing thickness of the current sheet in turn results in an infinitely broad spectral distribution for the generated radiation field.  But, given that it is created by the coordinated motion of aggregates of sub-luminally moving particles, a super-luminally moving source is necessarily volume-distributed: it can neither be point-like nor be distributed over a line or a surface~\citep{BolotovskiiBM:VaveaD, GinzburgVL:vaveaa, BolotovskiiBM:Radbcm}.  

In a physically more realistic model of the magnetosphere, where the processes that occur on plasma scales are taken into account, the current sheet would have a non-zero thickness and the divergence in question would not occur.  The singularity that arises from overlooking the finite width of the current sheet is removed in~\citet{Ardavan2021} by setting a lower limit, $c/(\kappa_{\rm u}\omega)$, on the wavelength of fluctuations associated with the microstructure of the pulse profile and treating $\kappa_{\rm u}$ as a free parameter.  The thickness of the current sheet is dictated by microphysical processes that are not well understood: the standard Harris solution of the Vlasov-Maxwell equations that is commonly used in analysing a current sheet is not applicable in the present case because the current sheet in the magnetosphere of a neutron star moves faster than light and so has no rest frame.  The upper limit on the frequency of the radiation is, in the present case, proportional to $\kappa_{\rm u}$, with a proportionality factor that is larger the closer are the turning points of the phase function shown in Fig.~\ref{FH6} to each other.  It turns out that the time-domain description of the radiation field, for large $\kappa_{\rm u}$, can be derived explicitly~\citep[][Sect. 4.7]{Ardavan2021}.

This radiation field consists of highly focused pulses whose salient features (brightness temperature, polarisation, spectrum and profile with microstructure and with a phase lag between the radio and gamma-ray peaks) are strikingly similar to those of the emission received from pulsars~\citep[see Sect. 5 and Figs.~9--16 of][]{Ardavan2021}.  That the flux density of the high-frequency component of this radiation decays non-spherically in certain latitudinal directions is moreover supported by the observational data on gamma-ray pulsars: the results of testing the hypothesis of independence of luminosities and distances of gamma-ray pulsars by means of the Efron--Petrosian statistic show that the data in the \textit{Fermi} Large Area Telescope (\textit{Fermi}-LAT) 12-Year Catalog are consistent with the dependence $S\propto D^{-3/2}$ of the flux densities $S$ of these pulsars on their distances $D$ at substantially higher levels of significance than they are with the dependence $S\propto D^{-2}$~\citep[see][]{Ardavan2023}.  

\begin{figure*}
\centerline{\includegraphics[width=16.7cm]{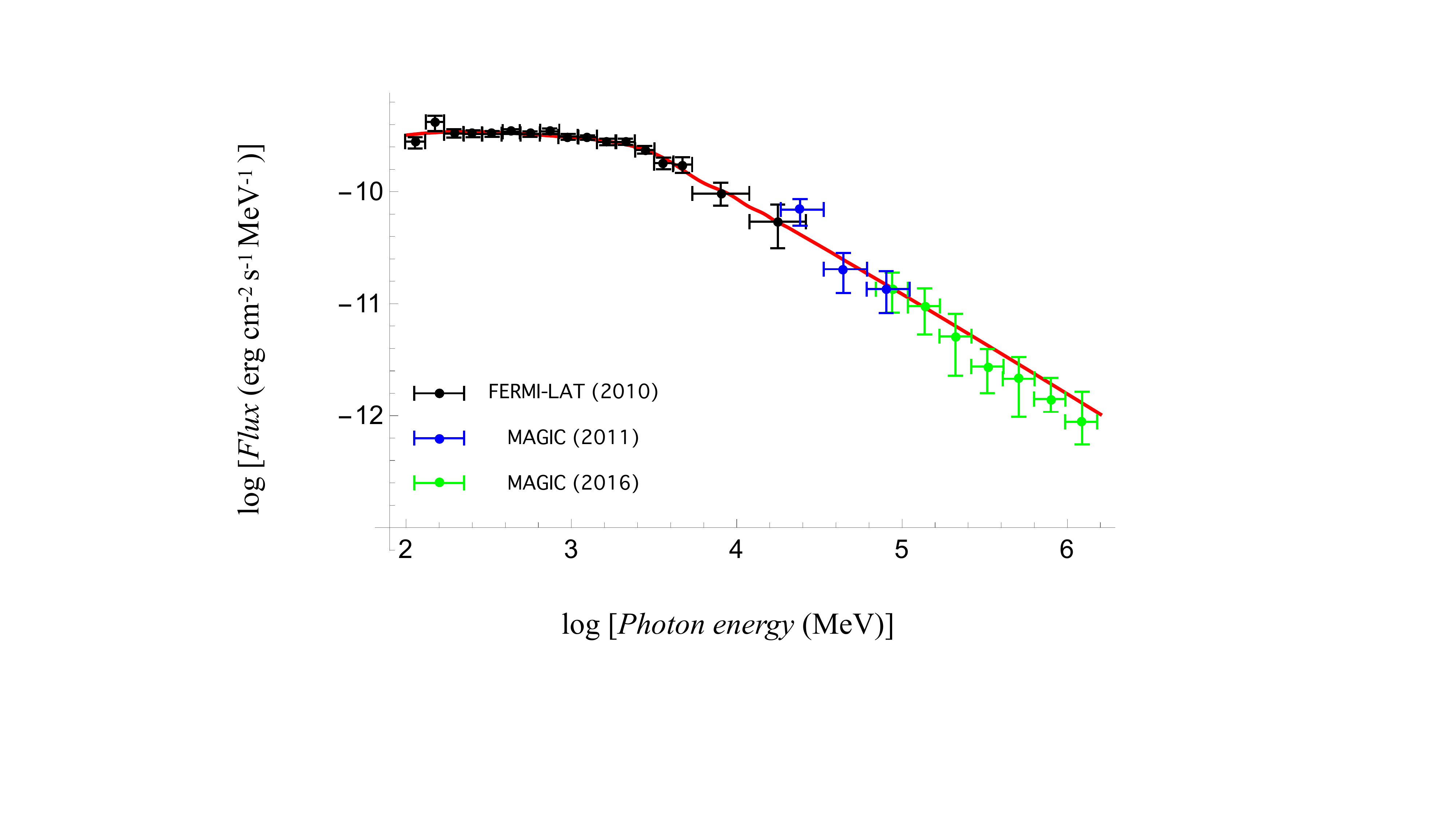}}
\caption{Phase-averaged gamma-ray spectrum of the Crab Pulsar.  The \textit{Fermi}-LAT data points below $25$ GeV (black) are those reported in~\citet{Abdo2010}.  The MAGIC data points between $25$ and $100$ GeV (blue) are those reported in~\citet{Aleksic2011}.  The MAGIC data points beyond $100$ Gev (green) are extracted (via the procedure described in Sect.~\ref{sec:fit}) from those reported by~\citet{Ansoldi2016}.  The red curve is a plot of the spectral distribution function described by Eq.~(\ref{E4}) for the parameters given in Eq.~(\ref{E9}).}
\label{CSF1}
\end{figure*}

\section{Spectral distribution function of tightly focused caustics}
\label{sec:spectrum}

The frequency spectrum of the radiation that is generated by the super-luminally moving current sheet in the magnetosphere of a non-aligned neutron star was presented, in its general form, in~\citet[][Sect.~5.3]{Ardavan2021}.  Here we derive the spectral distribution function of the most tightly focused component of this radiation from the general expression given in Eq. (177) of that paper.  

In a case where the magnitudes of the vectors denoted by $\boldsymbol{\cal P}_l$ and $\boldsymbol{\cal Q}_l$ in Eq. (177) of~\citet{Ardavan2021} are appreciably larger than those of their counterparts, ${\bar{\boldsymbol{\cal P}}}_l$ and ${\bar{\boldsymbol{\cal Q}}}_l$, and the dominant contribution towards the flux density $S_\nu$ of the radiation at the observation point is made by only one of the two terms corresponding to $l=1$ and $l=2$, Eq. (177) of~\citet{Ardavan2021} can be written as
\begin{equation}
S_\nu=\kappa_0\, k^{-2/3}\left\vert\boldsymbol{\cal P}\, {\rm Ai}(-k^{2/3}\sigma^2)-{\rm i}k^{-1/3}\boldsymbol{\cal Q}\,{\rm Ai}^\prime(-k^{2/3}\sigma^2)\right\vert^2,
\label{E1}
\end{equation}
where Ai and ${\rm Ai}^\prime$ are the Airy function and the derivative of the Airy function with respect to its argument, respectively, $k=2\pi\nu/\omega$ is the frequency $\nu$ of the radiation in units of the rotation frequency $\omega/2\pi$ of the central neutron star, and the positive scalars $\kappa_0$ and $\sigma$ and the complex vectors $\boldsymbol{\cal P}$ and $\boldsymbol{\cal Q}$ are determined by the characteristics of the current sheet.  The above spectrum is emblematic of any radiation that entails caustics~\citep[see][]{Stamnes1986}.

Evaluation of the right-hand side of Eq.~(\ref{E1}) results in
\begin{eqnarray}
S_\nu&=&\kappa_1\, k^{-2/3}\left[{\rm Ai}^2(-k^{2/3}\sigma^2)+\zeta_1^2k^{-2/3}{{\rm Ai}^\prime}^2(-k^{2/3}\sigma^2)\right.\nonumber\\*
&&\left.+2\zeta_1\cos\beta\, k^{-1/3}{\rm Ai}(-k^{2/3}\sigma^2){\rm Ai}^\prime(-k^{2/3}\sigma^2)\right],
\label{E2}
\end{eqnarray}
where 
\begin{equation}
\kappa_1=\vert\boldsymbol{\cal P}\vert^2\kappa_0,\qquad \zeta_1=\frac{\vert{\boldsymbol{\cal Q}}\vert}{\vert{\boldsymbol{\cal P}}\vert},\qquad\cos\beta=\frac{\Im(\boldsymbol{\cal Q}\cdot\boldsymbol{\cal P}^*)}{\vert\boldsymbol{\cal Q}\vert \vert\boldsymbol{\cal P}\vert},
\label{E3}
\end{equation}
and $\Im{}$ and $*$ denote an imaginary part and the complex conjugate, respectively.  The variable $\sigma$ has a vanishingly small value at those privileged latitudes (relative to the star's spin axis) where the high-frequency radiation is observable~\citep[][Sect.~4.5]{Ardavan2021}.  For the purposes of the analysis in Sect.~\ref{sec:fit}, we may therefore replace $\boldsymbol{\cal P}$ and $\boldsymbol{\cal Q}$ by their limiting values for $k\gg1$ and $\sigma\ll1$ and treat the quantities $\kappa_1$, $\zeta_1$ and $\beta$ that appear in Eq.~(\ref{E2}) as constant parameters.

To take account of the fact that the variable $\sigma$ assumes a non-zero range of values (bordering on $\sigma=0$) across the latitudinal width of a high-frequency radiation beam~\citep[][Sect.~3]{Ardavan2022a}, we must integrate $S_\nu$ with respect to $\sigma$ over a finite interval $0\le\sigma\le\sigma_0$ with $\sigma_0\ll1$.  Performing the integration of the Airy functions in Eq.~(\ref{E2}) with respect to $\sigma$ by means of Mathematica, we thus obtain 
\begin{eqnarray}
{\cal S}_\nu&=&\int_0^{\sigma_0}S_\nu\,{\rm d}\sigma\nonumber\\*
&=& \kappa\left[f_1(\chi)+\frac{\zeta^2}{4\sqrt{3}}f_2(\chi)-\frac{\zeta\cos\beta}{2\sqrt{3}}f_3(\chi)\right],
\label{E4}
\end{eqnarray}
where
\begin{equation}
\kappa=\left(\frac{{\sigma_0}}{\sqrt{3\pi}}\right)^3\kappa_1,\qquad \zeta=\sigma_0 \zeta_1,\qquad\chi=\frac{2}{3}{\sigma_0}^3k,
\label{E5}
\end{equation}
\begin{eqnarray}
f_1(\chi)&=&3\Gamma\left(\frac{7}{6}\right)\chi^{-2/3}{}_2F_3
\left(\begin{matrix}
1/6&1/6&{}\\
1/3&2/3&7/6
\end{matrix}
;-\chi^2
\right)\nonumber\\
&&+\pi^{1/2}{}_2F_3
\left(\begin{matrix}
1/2&1/2&{}\\
2/3&4/3&3/2
\end{matrix}
\,;-\chi^2
\right)\nonumber\\
&&+\frac{9}{20}\Gamma\left(\frac{5}{6}\right)\chi^{2/3}{}_2F_3
\left(\begin{matrix}
5/6&5/6&{}\\
4/3&5/3&11/6
\end{matrix}
\,;-\chi^2
\right),
\label{E6}
\end{eqnarray}
\begin{equation}
f_2(\chi)=\chi^{-4/3}\,{}_{24}G^{31}\left(-\chi^{2/3},\frac{1}{3}\left\vert\,\begin{matrix}
5/6&7/6&{}&{}\\
0&2/3&4/3&-1/6
\end{matrix}\right),\right.
\label{E7}
\end{equation}
\begin{equation}
f_3(\chi)=\chi^{-1}\,{}_{24}G^{31}\left(-\chi^{2/3},\frac{1}{3}\left\vert\,\begin{matrix}
5/6&1/2&{}&{}\\
0&1/3&2/3&-1/6
\end{matrix}\right),\right.
\label{E8}
\end{equation}
and ${}_2F_3$ and ${}_{24}G^{31}$ are respectively the generalised hypergeometric function~\citep[see][]{Olver} and the generalised Meijer G-function.\footnote{\url{https://mathworld.wolfram.com/Meijer G-Function.html}}  The variable $\chi$ that appears in the above expressions is related to the frequency $\nu$ of the radiation via $\chi=4\pi{\sigma_0}^3\nu/(3\omega)$.

The scale and shape of the spectral distribution function ${\cal S}_\nu(\nu)$ given in Eq.~(\ref{E4}) depend on the four parameters $\kappa$, $a$, $\beta$ and $\sigma_0$: parameters whose values are dictated by the characteristics of the magnetospheric current sheet.  The parameters $\kappa$ and $\sigma_0$ respectively determine the scale of the flux density and the breadth of the spectral distribution, while the parameters $\zeta$ and $\beta$ determine the shape of this distribution.

\section{Fitting the derived distribution function to the data on the gamma-ray spectrum of the Crab Pulsar}
\label{sec:fit}

The phase-averaged flux density of the gamma-rays received from the Crab Pulsar is plotted (in units of erg cm${}^{-2}$ s${}^{-1}$ MeV${}^{-1}$) versus photon energy (in units of MeV) in Fig.~\ref{CSF1}.  The data points below $25$ GeV in this figure are the ones detected by \textit{Fermi}-LAT, which were reported in~\citet{Abdo2010}.  The three data points between $25$ and $100$ GeV are those measured by the Major Atmospheric Gamma Imaging Cherenkov (MAGIC) telescope and were reported in~\citet{Aleksic2011}.  

Having been derived by a Fourier decomposition of the pulse profile with respect to time, Eq.~(\ref{E1}) and hence the expression for ${\cal S}_\nu$ in Eq.~(\ref{E4}) describe a spectral distribution that is averaged over the whole pulse period~\citep[see][]{Ardavan2021}.  On the other hand, the latest data by MAGIC on photon energies exceeding $100$ GeV are reported by~\citet{Ansoldi2016} for the main peak and the interpulse of the Crab Pulsar separately.  To estimate the phase-averaged fluxes at the five photon energies for which the fluxes associated with both the main peak and the interpulse are given in Table 3 of~\citet{Ansoldi2016}, we have here summed the contributions from these two components of the pulse profile.  Given that the reported upper limits on the contribution of the main pulse towards the fluxes at $781$ and $1211$ GeV are comparable to the observational errors in the contribution of the interpulse to the fluxes at these energies~\citep[see Table 3 of][]{Ansoldi2016}, we have equated the phase-averaged fluxes at $781$ and $1211$ GeV to the fluxes of the interpulse and have incorporated the upper limits on the contributions of the main pulse towards the phase-averaged fluxes at these two photon energies in the vertical error bars of the last two data points.

\begin{figure*}
\centerline{\includegraphics[width=19cm]{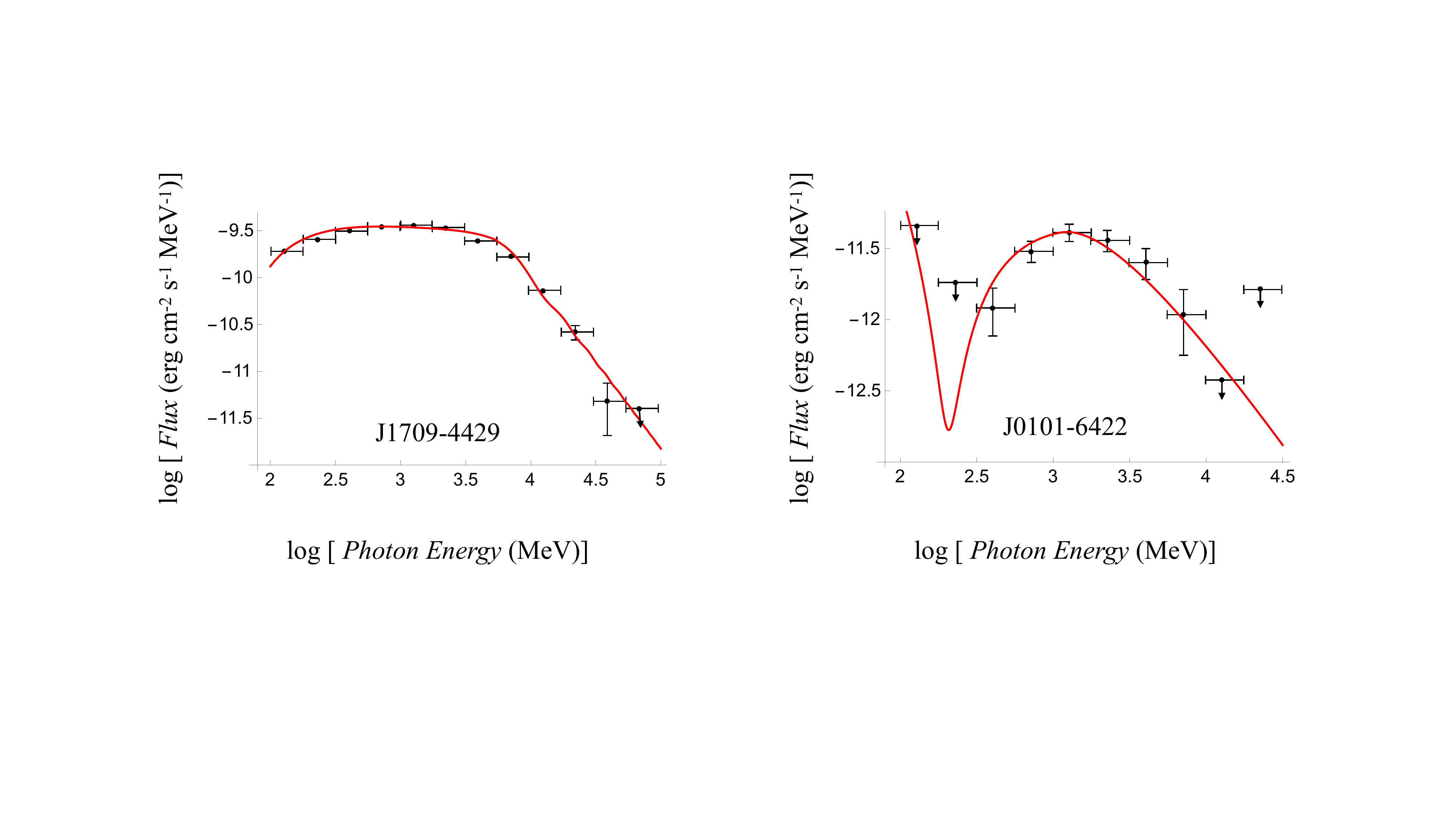}}
\caption{Phase-averaged gamma-ray spectra of PSR J1709-4429 and PSR J0101-6422.  The data are from the Second \textit{Fermi}-LAT Catalog of Gamma-ray Pulsars~\citep{Abdo2013}.  The red curves are plots of the spectral distribution function described by Eq.~(\ref{E4}) for the parameters given in Eqs.~(\ref{E10}) and (\ref{E11}).}
\label{CSF3}
\end{figure*}

The data point with the highest photon energy ($81$ GeV) in the dataset reported by~\citet{Aleksic2011} and the data point with the lowest photon energy ($87$ GeV) in the dataset reported by~\citet{Ansoldi2016} correspond to the same photon energy to within the limits of their horizontal error bars.  Hence, any discrepancy between the reported fluxes of these two data points must be due to the difference partly in the phase intervals over which these two group of authors have made their measurements and partly in the energy-dependent systematic uncertainties associated with their experiments~\citep[see][]{Aleksic2012}.  Accordingly, we have removed the discrepancy between the phase-averaged fluxes of the lowest-energy data point extracted from~\citet{Ansoldi2016} and the highest-energy data point given in~\citet{Aleksic2011} by multiplying the former by the factor $9.77$.  Once the fluxes of each of the remaining six data points extracted from~\citet{Ansoldi2016} are multiplied by the same factor, these points are collectively raised to the level shown in Fig.~\ref{CSF1}.  

The curve in Fig.~\ref{CSF1} depicts the spectral distribution function ${\cal S}_\nu$ that is described by Eq.~(\ref{E4}) for the following values of its free parameters:
\begin{eqnarray}
\kappa&=&7.94\times10^{-11}\quad{\rm erg}\,{\rm cm}^{-2}\,{\rm s}^{-1}\,{\rm MeV}^{-1},\nonumber\\*
\chi&=&5.62\times10^{-4}\times(h\nu)_{\rm MeV},\nonumber\\*
\sigma_0&=&4.72\times10^{-8},\quad \zeta=0.4\quad{\rm and}\quad\beta=0.3,
\label{E9}
\end{eqnarray}
in which $h$ stands for the Planck constant and $2\pi/\omega$ has been set equal to the period of the Crab Pulsar.  As the photon energy $h\nu$ increases beyond $1.2$ TeV, the slope of this curve continues to decrease at the relatively slow rate in which it decreases past $50$ GeV.  Thus, the upper limits given by~\citet{Ansoldi2016} for the flux density of gamma-rays whose energies exceed $1.2$ TeV all lie above the continuation of the curve shown in Fig.~\ref{CSF1}. 

The relationships between the above parameters of the observed gamma-ray spectrum of the Crab Pulsar and the physical characteristics of the source of this spectrum are derived in Appendix~\ref{App}.  The values of the parameters in Eq.~(\ref{E9}) imply (i) that the angle between the magnetic and spin axes of the Crab Pulsar has the value $\alpha=75.5^\circ$, (ii) that the scale factor of the electric current density associated with the magnetospheric current sheet of this pulsar is given by $j_0=2.10\times10^3$ amp m${}^{-2}$, and (iii) that the privileged direction along which the high-frequency radiation from this pulsar is observed has the colatitude $\theta_{PS}=132.9^\circ$ with respect to the spin axis of its central neutron star (see Appendix~\ref{App}). 

The accuracy with which the free parameters of the fitted spectra are specified in this and the next section reflects that with which the characteristics of their underlying neutron stars (period, magnetic field strength, distance, etc.) are known.  It should be borne in mind, however, that the semi-analytic description of the magnetospheric structure of a pulsar given in~\citet[][Sect. 2]{Ardavan2021}, on which the present analysis is based, might not be as accurate.

\section{Examples of spectra of other gamma-ray pulsars}
\label{sec:examples}

To emphasise that the function given by Eq.~(\ref{E4}) describes the spectral distribution of the high-frequency emission from any non-aligned neutron star, we have applied the procedures outlined in Sect.~\ref{sec:fit} and Appendix~\ref{App} also to the spectra of two other pulsars for which the available datasets are sufficiently large.  Figure~\ref{CSF3} shows the results of fitting this spectral distribution function to the data on the spectra of PSR J1709-4429 and PSR J0101-6422 reported by~\citet{Abdo2013}.

In the case of PSR J1709-4429, the free parameters of ${\cal S}_\nu$ have the values
\begin{eqnarray}
\kappa&=&8.91\times10^{-11}\quad{\rm erg}\,{\rm cm}^{-2}\,{\rm s}^{-1}\,{\rm MeV}^{-1},\nonumber\\*
\chi&=&3.16\times10^{-4}\times(h\nu)_{\rm MeV},\nonumber\\*
\sigma_0&=&2.68\times10^{-8},\quad \zeta=0.4\quad{\rm and}\quad\beta=0
\label{E10}
\end{eqnarray}
for photon energies below $\log(h\nu)_{\rm MeV}=3.74$.  At higher photon energies, the factor $k^{-2/3}$ that multiplies $\kappa_0$ in the expression for $S_\nu$ in Eq.~(\ref{E1}) should be replaced by $k^{-5/3}$ (see Appendix~\ref{App}), as a result of which the values of $\kappa$ and $\zeta$ change to $4.47\times10^{-7}$ and $1.5$, respectively.  The parameters in Eq.~(\ref{E10}) imply (i) that the inclination angle of PSR J1709-4429 has the value $\alpha=58^\circ$, (ii) that the scale factor of the electric current density associated with the magnetospheric current sheet of this pulsar is given by $j_0=2.36\times10^3$ amp m${}^{-2}$, and (iii) that the privileged direction along which the high-frequency radiation from this pulsar is observed has the colatitude $\theta_{PS}=85.4^\circ$ with respect to the spin axis of its central neutron star (see Appendix~\ref{App}).

In the case of PSR J0101-6422, the free parameters of ${\cal S}_\nu$ have the values
\begin{eqnarray}
\kappa&=&9.44\times10^{-22}\quad{\rm erg}\,{\rm cm}^{-2}\,{\rm s}^{-1}\,{\rm MeV}^{-1},\nonumber\\*
\chi&=&1.26\times10^{-18}\times(h\nu)_{\rm MeV},\nonumber\\*
\sigma_0&=&1.49\times10^{-12},\quad \zeta=10^{-5}\quad{\rm and}\quad\beta=0.05
\label{E11}
\end{eqnarray}
for photon energies below $\log(h\nu)_{\rm MeV}=3.05$.  In this case, too, the factor $k^{-2/3}$ that multiplies $\kappa_0$ in the expression for $S_\nu$ should be replaced by $k^{-5/3}$ at higher energies, as a result of which the values of $\kappa$, $\chi,$ and $\sigma_0$ change to $1.78\times10^{-18}$, $5.62\times10^{-19}$, and $1.11\times10^{-12}$, respectively (see Appendix~\ref{App}).  The parameters in Eq.~(\ref{E11}) imply (i) that the inclination angle of PSR J0101-6422 has the value $\alpha=0.021^\circ$, (ii) that the scale factor of the electric current density associated with the magnetospheric current sheet of this pulsar is given by $j_0=7.35\times10^2$ amp m${}^{-2}$, and (iii) that the privileged direction along which the high-frequency radiation from this pulsar is observed has the colatitude $\theta_{PS}=0.021^\circ$ with respect to the spin axis of its central neutron star (i.e. it\ has the same value as $\alpha$; see Appendix~\ref{App}).  The low value of the inclination angle for which the spectral distribution function in Eq.~(\ref{E4}) fits the spectrum of this pulsar stems, according to Eq.~(\ref{A12}) and Fig.~\ref{CSF2}, from the low value of the magnetic field ($1.17\times10^8$ Gauss) at the magnetic pole of its central neutron star (see the ATNF Catalogue; \citealt{YMW16}).  This example illustrates that for any non-zero inclination angle, no matter how small, the structure of the current sheet shown in Fig.~\ref{FH0} (and hence the electromagnetic field generated by this sheet) is manifestly different from that of an equatorially confined planar one (i.e.\ different from the structure of the current sheet for an aligned rotator).
 
\section{Concluding remarks}
\label{sec:conclusion}
In the current literature on the gamma-ray spectrum of the Crab Pulsar~\citep[see the review articles][and the references therein]{Zanin2017,Amato2021}, this spectrum is parametrised partly (between $0.1$ and $100$ GeV) by a power-law function with an exponential cutoff and partly (between $100$ GeV and $1$ TeV) by a pure power law with another exponent.  The part of the emission between $0.1$ and $100$ GeV is commonly attributed to synchrotron, curvature, or synchro-curvature processes.  It is generally acknowledged, however, that the very high energy component of the emission is unlikely to be generated by such processes.   The emission at higher than a few hundred GeV is thought to be produced either by inverse Compton scattering of lower energy photons or as a consequence of magnetic reconnection in the magnetospheric current sheet.

We have here presented a fit to the entire gamma-ray spectrum of the Crab Pulsar, which, in contrast to the disparate phenomenological ones currently considered in the literature on the subject, is a consequence simply of the first principles that govern the magnetospheric structure of a neutron star and the emission of electromagnetic waves (see Sect.~\ref{sec:heuristic}).  Not only do the results presented here provide an explanation for both the shape and the exceptional breadth of the Crab Pulsar's spectrum in terms of a single emission mechanism, but they also shed light on certain physical characteristics of this pulsar's magnetosphere: the angle between the magnetic and spin axes of its central neutron star, the scale factor of the electric current density that is associated with its current sheet, and the privileged latitudinal direction (relative to the spin axis) in which it is observed (Sect.~\ref{sec:fit}).

The relevance of the spectral distribution function given in Eq.~(\ref{E4}) is of course not limited to the spectrum of the Crab Pulsar; this function describes the spectral distribution of the high-frequency emission from any non-aligned neutron star.  We have illustrated this by analysing, in addition, data on the spectra of two other gamma-ray pulsars, PSR J0101-6422 and PSR J1709-4429, for which the free parameters in Eq.~(\ref{E4}), and thus the physical characteristics of the current sheets that underlie their emissions, have widely different values (Sect.~\ref{sec:examples}).

\bibliographystyle{aa}
\bibliography{CrabSpec.bib}

\newpage

\begin{appendix}
\section{In which parameters of the observed spectra are related to the physical characteristics of their sources}
\label{App}
\setcounter{equation}{0}

From the values of the parameters $\kappa$, $\chi,$ and $\sigma_0$ of the observed gamma-ray spectra of pulsars, we can derive the angle $\alpha$ between the magnetic and spin axes of their central neutron stars, the colatitude $\theta_P$ along which they are observed and the scale factor $j_0$ of the electric current density of the current sheet in their magnetospheres.

According to Eqs.~(177), (138), (144), and (136) of \citet{Ardavan2021}, the factor $\kappa$ that appears in Eq.~(\ref{E4}) is related to the characteristics of the source of the observed radiation via Eqs.~(\ref{E5}) and (\ref{E3}),
\begin{equation}
\kappa_0=4.15\times10^{18} w_1^2 w_3^2 {\hat B}_0^2 d^4 D^{-2} {\hat P}^{-1}\quad {\rm Jy},
\label{A1}
\end{equation}
and
\begin{equation}
\vert\boldsymbol{\cal P}\vert=\frac{1}{3}b\left(\frac{2\sigma_{21}}{\partial^2 f_{2C}/\partial\tau^2|_{\tau=\tau_{2{\rm min}}}}\right)^{1/2}\vert{\tilde{\bf P}}_2\vert,
\label{A2}
\end{equation}
in which
\begin{equation}
w_1=\left\vert1-2\alpha/\pi\right\vert,\qquad w_3=1+0.2\sin^2\alpha,
\label{A3}
\end{equation}
\begin{equation}
B_0=10^{12}{\hat B}_0\,\,{\rm Gauss},\quad  R_P=D\,\,{\rm kpc}=3.085\times10^{21}D\,\,{\rm cm},
\label{A4}
\end{equation}\begin{equation}
\omega=10^2{\hat P}^{-1}\,\,{\rm rad}/{\rm s},\qquad r_{s0}=10^6 d\,\,{\rm cm},
\label{A5}
\end{equation}
\begin{eqnarray}
f_{2C}&=&({\hat r}_P^2{\hat r}_{sC}^2\sin^2\theta-1)^{1/2}-{\hat R}_P+{\hat r}_{sC}\nonumber\\*
&&-\arccos\left({\hat r}_P^{-1}{\hat r}_{sC}^{-1}\csc\theta\right)\nonumber\\*
&&-\arccos(\cot\alpha\cot\theta)+\varphi_P-{\hat r}_{s0}-2\pi,
\label{A6}
\end{eqnarray}
\begin{eqnarray}
a&=&({\hat r}_P^2{\hat r}_{sC}^2\sin^2\theta-1)\left[\frac{({\hat r}_P^2{\hat r}_{sC}^2\sin^2\theta-1)^{1/2}+{\hat r}_{sC}}{{\hat r}_{sC}({\hat r}_P^2-1)^{1/2}({\hat R}_P^2\sin^2\theta-1)^{1/2}}\right.\nonumber\\*
&&\left.-\frac{1}{({\hat r}_P^2{\hat r}_{sC}^2\sin^2\theta-1)^{1/2}}\right],
\label{A7}
\end{eqnarray}
\begin{equation}
b=\frac{{\hat r}_P^2{\hat r}_{sC}^2\sin^2\theta-1}{{\hat r}_{sC}({\hat r}_P^2-1)^{1/2}({\hat R}_P^2\sin^2\theta-1)^{1/2}},
\label{A8}
\end{equation}
\begin{equation}
{\hat r}_{sC}=\frac{({\hat r}_P^2-1)^{1/2}({\hat R}_P^2\sin^2\theta-1)^{1/2}-{\hat z}_P\cos\theta}{{\hat r}_P^2\sin^2\theta-1},
\label{A9}
\end{equation}
\begin{equation}
\sigma_{21}=[{\textstyle\frac{3}{4}}(f_{2C}\vert_{\tau=\tau_{2{\rm max}}}-f_{2C}\vert_{\tau=\tau_{2{\rm min}}})]^{1/3},
\label{A10}
\end{equation}
\begin{eqnarray}
\vert{\tilde{\bf P}}_2\vert &=&\cos\alpha[{\hat r}_{sC}^2(1+\cos^2\theta_P-2\cos\theta\cos\theta_P)+\tan^2\alpha\nonumber\\*
&&-\cot^2\theta+2{\hat r}_{sC}\sin\theta\cos\theta_P(\tan^2\alpha-\cot^2\theta)^{1/2}]^{1/2},\nonumber\\*
\label{A11}
\end{eqnarray}
and $\theta=\arccos(\sin\alpha\cos\tau)$ (see Eqs.~(7), (9), (174), (175), (93)--(95), (97), and (88) of~\citealt{Ardavan2021}).  In these expressions, $\alpha$, $B_0$, $r_{s0}$, respectively denote the inclination angle of the central neutron star, the magnitude of the star's magnetic field at its magnetic pole, the radius of the star, and the spherical polar coordinates of the observation point, $P,$ in a frame whose centre and $z$-axis coincide with the centre and spin axis of the star.  The caret on $R_P$ and $r_{s0}$ (and $r_P=R_P\sin\theta_P$, $z_P=R_P\cos\theta_P$) is used here to designate a variable that is rendered dimensionless by being measured in units of the light-cylinder radius $c/\omega$.  The variables $\tau_{2{\rm min}}$ and $\tau_{2{\rm max}}$ stand for the minimum and maximum of the function $f_{2C}$.  To be specific, we have set the integer $l$ that appears in the expression for $f_{lC}$ equal to $2$.  This choice does not affect the results that follow since, as pointed out in Sect.~4.4 of~\citet{Ardavan2021}, $f_{2C}$ is transformed into $f_{1C}$ by the changes $\theta_P\to\pi-\theta_P$, $\varphi_P\to\pi+\varphi_P$ and $\theta\to\pi-\theta$.  We have also corrected a typographical error in Eq.~(177) of~\citet{Ardavan2021}.

The expression for $\vert{\tilde{\bf P}}_2\vert$ in Eq.~(\ref{A11}) is derived from that for ${\bf P}_2$ given by Eqs.~(98), (80), (78) and (62) of~\citet{Ardavan2021}.  In this derivation, we have set the observation point on the cusp locus of the bifurcation surface where ${\hat r}_s={\hat r}_{sC}$, have approximated $(p_1,p_2,p_3)$ by its far-field value $2^{1/3}{\hat R}_P^{-1}({\hat R}_P^{-1}, -1, 1)$ and have let ${\bf P}_2={\hat R}_P^{-1}{\tilde{\bf P}}_2$.  The factor ${\hat R}_P^{-2}$ that would have otherwise appeared in the resulting expression for $\vert\boldsymbol{\cal P}\vert^2$ (see Eq.~(\ref{A2})) is thus incorporated in the factor $\kappa_0$ (see Eq.~(\ref{A1})).

Given that $1$ Jy $=2.418\times10^{-3}$ erg cm${}^{-2}$ s${}^{-1}$ MeV${}^{-1}$, Eqs.~(\ref{E5}), (\ref{E3}), (\ref{A1}), and (\ref{A2}) jointly yield
\begin{equation}
{\hat\kappa}_{\rm th}=2.60\times10^{-14}D^2{\hat P}{\hat B}_0^{-2}d^{-4}\sigma_0^{-3}\kappa
\label{A12}
,\end{equation}
in which
\begin{equation}
{\hat\kappa}_{\rm th}=w_1^2 w_3^2 b^2\left(\frac{2 \sigma_{21}} {\partial^2 f_{2C}/\partial\tau^2|_{\tau=\tau_{2{\rm min}}}}\right)\vert{\tilde{\bf P}}_2\vert^2,
\label{A13}
\end{equation} and $\kappa$ is in units of erg cm${}^{-2}$ s${}^{-1}$ MeV${}^{-1}$.  While the right-hand side of Eq.~(\ref{A12}) only contains the observed parameters of the pulsar and its emission, the value of ${\hat\kappa}_{\rm th}$, which appears on the left-hand side of this equation, is determined, according to Eq.~(\ref{A13}), by the physical characteristics of the current sheet in the pulsar's magnetosphere.

\begin{figure}
\centerline{\includegraphics[width=10cm]{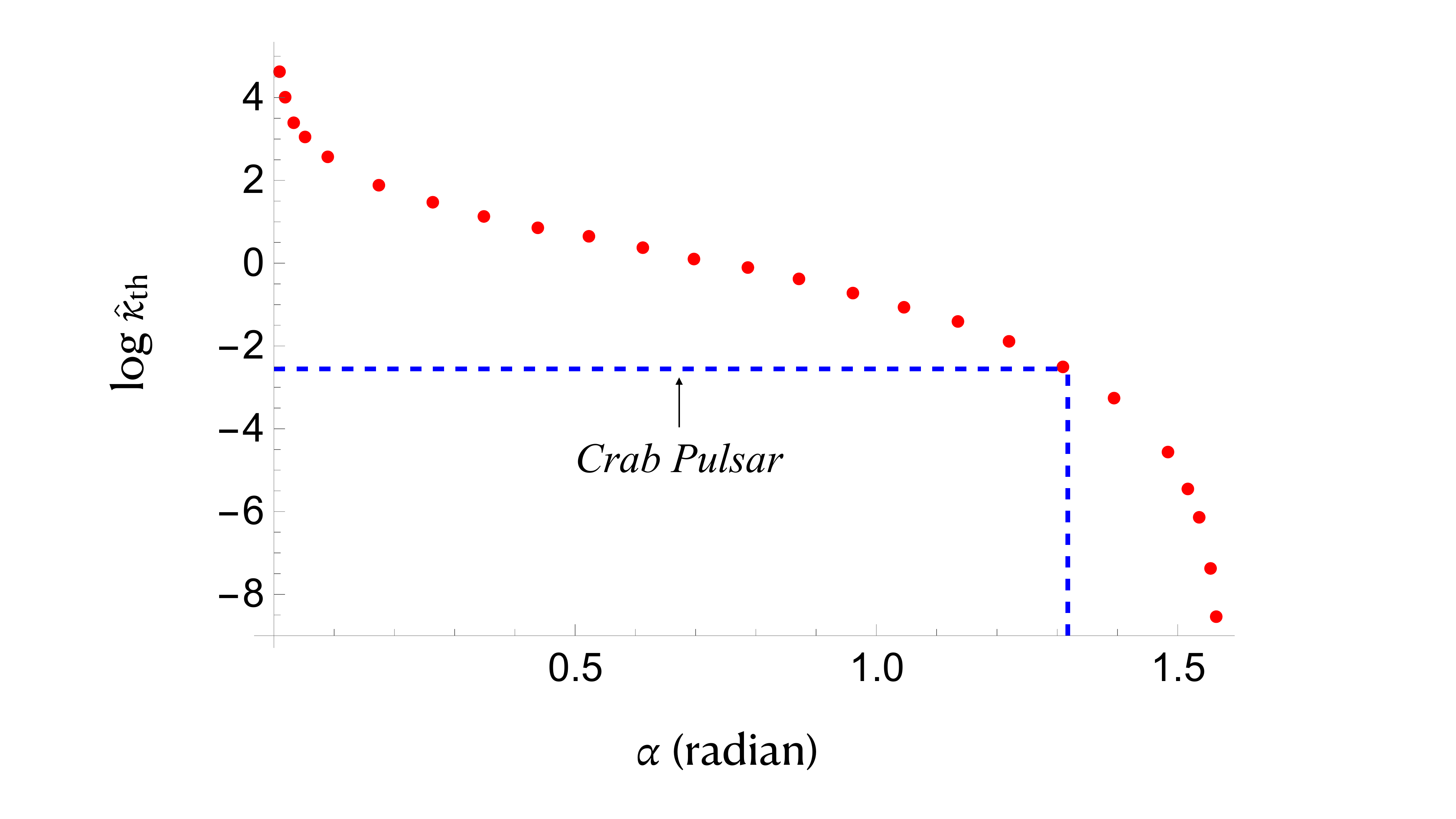}}
\caption{Dependence on the inclination angle, $\alpha,$ of the theoretical expression for ${\hat\kappa}_{\rm th}$, given by Eq.~(\ref{A13}), for $\theta_P\to\theta_{PS}$, $\sigma_{21}\to0,$ and $\tau\to\tau_{2{\rm min}}\to\tau_{2{\rm max}}$, shown by the red points.  The theoretical value of ${\hat\kappa}_{\rm th}$ ranges from infinity to zero over the interval $0<\alpha\le\pi/2$.  The horizontal dashed line (in blue) shows the value of ${\hat\kappa}_{\rm th}$, given by Eq.~(\ref{A12}), for the parameters of the Crab Pulsar derived from observations.  As indicated by the vertical dashed line (in blue), these two values of ${\hat\kappa}_{\rm th}$ match for an $\alpha=1.32$ radian (i.e.\ $\alpha=75.5^\circ$).}
\label{CSF2}
\end{figure}

For certain values of $\theta_P$, here denoted by $\theta_{PS}$, the function $f_{2C}(\tau)$ has an inflection point (see Sect.~\ref{sec:heuristic}).  For any given inclination angle $\alpha$, the position $\tau_{2S}$ of this inflection point and the colatitude $\theta_{PS}$ of the observation points for which $f_{2C}(\tau)$ has an inflection point follow from the solutions to the simultaneous equations $\partial f_{2C}/\partial\tau=0$ and $\partial^2 f_{2C}/\partial\tau^2=0$.  (Explicit expressions for the derivatives that appear in these equations can be found in Appendix A of~\citet{Ardavan2021}.)   For values of $\theta_P$ close to $\theta_{PS}$, the separation between the maximum $\tau=\tau_{2{\rm max}}$ and minimum $\tau=\tau_{2{\rm min}}$ of $f_{2C}$ and hence the value of $\sigma_{21}$ are correspondingly small.  Since the high-frequency radiation that arises from the current sheet is detectable only in the vicinity of the latitudinal direction $\theta_P=\theta_{PS}$, here we evaluate ${\hat\kappa}_{\rm th}$ at a given inclination angle $\alpha$ for $\sigma_{21}\to0$.  The ratio appearing inside the parentheses in Eq.~(\ref{A13}), which becomes indeterminate as $\sigma_{21}$ tends to zero, approaches a finite value.  Once $\vert{\tilde{\bf P}}_2\vert$ is also evaluated for $\tau=\tau_{2{\rm min}}$, which has the same value as $\tau_{2{\rm max}}$ in this limit, Eq.~(\ref{A13}) yields ${\hat\kappa}_{\rm th}$ as a function of $\alpha$.  The result is shown in Fig.~\ref{CSF2}.

According to Eq.~(\ref{A12}), on the other hand, ${\hat\kappa}_{\rm th}$ has the value $2.88\times10^{-3}$ for the parameters of the Crab Pulsar derived from observations, that is to say,\ for $\kappa=7.94\times10^{-11}$, $\sigma_0=4.72\times10^{-8}$ (see Eq.~(\ref{E9})), $D=2$, ${\hat P}=0.527$, ${\hat B}_0=3.79,$ and $d=1$ (see the ATNF Catalogue; \citealt{YMW16} and Eqs.~(\ref{E9}), (\ref{A4}) and (\ref{A5})).  As can be seen from Fig.~\ref{CSF2}, this observationally determined value of ${\hat\kappa}_{\rm th}(\alpha)$ matches the theoretically determined value of ${\hat\kappa}_{\rm th}(\alpha)$ given by Eq.~(\ref{A13}) for $\alpha=1.32$ radian (i.e.\ $\alpha=75.5^\circ$).  This value of the inclination angle in turn determines the scale factor $j_0$ of the electric current density: from Eq.~(22) of~\citet{Ardavan2021} and Eqs.~(\ref{A3})--(\ref{A5}) it follows that
\begin{equation}
j_0=1.29\times10^8w_1w_3{\hat B}_0{\hat P}^{-3}d^2\qquad{\rm statamp}\,\,  {\rm cm}^{-2}
\label{A14}
,\end{equation}
which has the value $6.31\times10^8$ statamp cm${}^{-2}$ (i.e.\ $2.10\times10^3$ amp m${}^{-2}$) for the Crab Pulsar.  Furthermore, the colatitude of the points along which the high-frequency radiation from this pulsar is detectable is given by $\theta_{PS}=132.9^\circ$, according to the solution to the simultaneous equations $\partial f_{2C}/\partial\tau=0$ and $\partial^2 f_{2C}/\partial\tau^2=0$.  

It should be noted that there is also a relationship between the parameter $\zeta$ in Eq.~(\ref{E9}) and the magnitude of the vector $\boldsymbol{\cal Q}$ in Eq.~(\ref{E3}).  However, the observationally determined value of $\vert\boldsymbol{\cal Q}\vert$ turns out to be appreciably larger than its theoretically determined value: a discrepancy that may stem from the approximate nature of the semi-analytic description of the magnetosphere used in Sect.~2 of~\citet{Ardavan2021}.

In the case of PSR J0101-6422 for which $\kappa=9.44\times10^{-22}$, $\sigma_0=1.49\times10^{-12}$ (see Eq.~(\ref{E10})),   $D=1.001$, ${\hat P}=4.09\times10^{-2}$, ${\hat B}_0=1.17\times10^{-4}$, and $d=1$ (see the ATNF Catalogue, ~\citealt{YMW16}, and Eqs.~(\ref{A4}) and (\ref{A5})), the above procedure yields $2.22\times10^7$ for the observationally determined value of ${\hat\kappa}_{\rm th}$, a value that matches the theoretically determined value of ${\hat\kappa}_{\rm th}$ for an $\alpha=3.67\times10^{-4}$ radian (i.e.\ $0.021^\circ$; see Fig.~\ref{CSF2}).  In this case, the value of  $\theta_{PS}$ is also $0.021^\circ$ and $j_0$ equals $2.21\times10^8$ statamp cm${}^{-2}$ or $7.35\times10^2$ amp m${}^{-2}$.  In contrast to the function ${\cal S}_\nu$ for the Crab Pulsar in which the term $\boldsymbol{\cal P}^{(0)}_2$ in the expression for $\boldsymbol{\cal P}_2$ in Eq.~(138) of~\citet{Ardavan2021} is dominant at all frequencies, in this case the term $\boldsymbol{\cal P}^{(2)}_2$ in that expression dominates over $\boldsymbol{\cal P}^{(0)}_2$ in the range of photon energies $\log(h\nu)_{\rm MeV}\gtrsim3.05$.  As a result, the factor $k^{-2/3}$ that multiplies $\kappa_0$ in Eq.~(\ref{E1}) changes to $k^{-5/3}$ at higher photon energies.

In the case of PSR 1709-4429 for which $\kappa=8.91\times10^{-11}$, $\sigma_0=2.68\times10^{-8}$ (see Eq.~(\ref{E11})) and  $D=2.6$, ${\hat P}=1.63$, ${\hat B}_0=3.12$ and $d=1$ (see the ATNF Catalogue, ~\citealt{YMW16}, and Eqs.~(\ref{A4}) and (\ref{A5})), the above procedure yields $1.36\times10^{-1}$ for the observationally determined value of ${\hat\kappa}_{\rm th}$, a value that matches the theoretically determined value of ${\hat\kappa}_{\rm th}$ for $\alpha=1.01$ radian (i.e.\ $58^\circ$; see Fig.~\ref{CSF2}).  In this case, the value of  $\theta_{PS}$ is $85.4^\circ$ and $j_0$ equals $7.09\times10^8$ statamp cm${}^{-2}$ or $2.36\times10^3$ amp m${}^{-2}$.  Here, too, as in the case of PSR J0101-6422, the term $\boldsymbol{\cal P}^{(2)}_2$ in the expression for $\boldsymbol{\cal P}_2$ in Eq.~(138) of~\citet{Ardavan2021} dominates over the term $\boldsymbol{\cal P}^{(0)}_2$ in the range of photon energies $\log(h\nu)_{\rm MeV}\gtrsim3.74$. 

\end{appendix}

\end{document}